\documentclass[12pt]{article}
\usepackage{amssymb}
\usepackage{amsmath}
\usepackage{diagbox}
\usepackage[ruled,vlined]{algorithm2e}

\numberwithin{equation}{section}
\def \bes{\begin{eqnarray}}
\def \ees{\end{eqnarray}}
\def \bns{\begin{eqnarray*}}
\def \ens{\end{eqnarray*}}

\newcommand{\be}{\begin{eqnarray}}
\newcommand{\ee}{\end{eqnarray}}
\newcommand{\non}{\nonumber}

\begin{document}

\begin{titlepage}
\strut\hfill UMTG--277
\vspace{.5in}
\begin{center}

\LARGE On the completeness of solutions of Bethe's equations\\
\vspace{1in}
\large Wenrui Hao \footnote{
Mathematical Biosciences Institute, The Ohio State University, 1735
Neil Avenue, Columbus, OH 43210 USA,
hao.50@osu.edu}, Rafael I. Nepomechie \footnote{Physics Department,
P.O. Box 248046, University of Miami, Coral Gables, FL 33124 USA,
nepomechie@physics.miami.edu}
and Andrew J. Sommese \footnote{
Department of Applied and Computational Mathematics and Statistics,
University of Notre Dame, 163 Hurley Hall, IN 46556 USA,
sommese@nd.edu}
\\[0.8in]
\end{center}

\vspace{.5in}

\begin{abstract}
    We consider the Bethe equations for the isotropic spin-1/2
    Heisenberg quantum spin chain with periodic boundary conditions.
    We formulate a conjecture for the number of solutions with
    pairwise distinct roots of these equations, in terms of numbers
    of so-called singular (or exceptional) solutions.  Using
    homotopy continuation methods, we find all such solutions of the Bethe
    equations for chains of length up to 14.  The numbers of these
    solutions are in perfect agreement with the conjecture.  We also
    discuss an indirect method of finding solutions of the Bethe
    equations by solving the Baxter T-Q equation.  We briefly comment
    on implications for thermodynamical computations based on the
    string hypothesis.
\end{abstract}

\end{titlepage}

\setcounter{footnote}{0}

\section{Introduction}\label{sec:intro}

The Heisenberg quantum spin chain is a one-dimensional array of $N$
quantum spin-1/2 spins with nearest-neighbor isotropic interactions
and periodic boundary conditions.  The Hamiltonian is given by
\be
H =  \frac{1}{4}\sum_{n=1}^{N}  \left(
\vec \sigma_{n} \cdot \vec \sigma_{n+1} - 1 \right)  \,, \qquad \vec
\sigma_{N+1} \equiv \vec \sigma_{1} \,,
\label{Hamiltonian}
\ee
where $\vec \sigma = (\sigma^{x}\,, \sigma^{y}\,, \sigma^{z})$ are the
usual $2 \times 2$ Pauli spin matrices, and $\vec \sigma_{n}$ denotes
the spin operators at site $n$.  The vector space describing this
system has dimension $2^{N}$, so the Hamiltonian is a $2^{N} \times
2^{N}$ matrix.  The basic problem is to determine the eigenvectors and
eigenvalues of this matrix, which grows in size exponentially with $N$.

The Heisenberg spin chain is of fundamental importance in theoretical
physics.  It is a model of (anti)ferromagnetism, which is realized
experimentally (e.g., KCuF${}_{3}$ and Sr${}_{2}$CuO${}_{3}$)
\cite{Mikeska:2004}.  It has many other applications and connections,
including conformal field theory \cite{Alcaraz:1987zr,
Affleck:1988mua}, ${\cal N}=4$ supersymmetric Yang-Mills theory and
string theory \cite{Beisert:2010jr}.  Moreover, it is the prototype of
so-called quantum integrable models: one-dimensional many-body quantum
systems that have many conserved quantities, and that therefore admit
exact solutions \cite{Baxter1982, Korepin:1993, Faddeev:1996iy,
Nepomechie:1998jf}.

An exact solution of the Heisenberg model was discovered in 1931 by Hans Bethe
\cite{Bethe:1931hc}.  In particular, the eigenvalues of the Hamiltonian
(\ref{Hamiltonian}) are given by
\be
E = - {1\over 2} \sum_{k=1}^{M} \frac{1}{\lambda_{k}^{2} + \frac{1}{4}} \,,
\label{energy}
\ee
where $\{\lambda_{1} \,, \ldots \,, \lambda_{M}\}$ satisfy Bethe's celebrated equations
\be
\left(\lambda_{k} + \frac{i}{2}\right)^{N}
\prod_{\scriptstyle{j \ne k}\atop \scriptstyle{j=1}}^M
(\lambda_{k} - \lambda_{j} - i)
= \left(\lambda_{k} - \frac{i}{2}\right)^{N}
\prod_{\scriptstyle{j \ne k}\atop \scriptstyle{j=1}}^M
(\lambda_{k} - \lambda_{j} + i) \,, \non \\
\qquad k = 1 \,, 2\,, \ldots \,, M \,, \qquad
M = 0\,, 1\,, \ldots\,,  \frac{N}{2} \,.
\label{BAE}
\ee
It is therefore customary to refer to the $\lambda_{k}$'s as ``Bethe
roots''. The solutions of other quantum integrable models entail
generalizations of these equations.

For given values of $N$ and $M$, the Bethe equations have various sets
of solutions.  Roughly speaking, for each such solution $\{\lambda_{1}
\,, \ldots \,, \lambda_{M}\}$, there is a corresponding energy level
(\ref{energy}) and eigenvector (the construction of which will be
sketched in Section \ref{sec:completeness}) of the Hamiltonian.
However, ever since the time of Bethe's remarkable discovery, the
nagging question of ``completeness'' has persisted: namely, whether
the Bethe equations have too many, too few, or just the right number
of solutions to account for all $2^{N}$ eigenstates of the
Hamiltonian.\footnote{Various completeness proofs have been proposed
for the case $N\rightarrow \infty$ \cite{Faddeev:1996iy, Bethe:1931hc,
Takahashi1971, Kirillov:1985}.  However, these proofs rely on the
so-called string hypothesis, which itself has not been proved.  See
also e.g. \cite{Langlands:1995, Langlands:1997, Fabricius:2000yx,
Baxter:2001sx, Mukhin:2009}.} The existence of so-called singular (or
exceptional) solutions of the Bethe equations makes the completeness
problem particularly confusing.\footnote{There has been considerable
discussion in the literature about such solutions e.g.
\cite{Avdeev:1985cx, Essler:1991wd, Siddharthan:1998, Noh:2000,
Beisert:2003xu, Beisert:2004hm, Hagemans:2007, Arutyunov:2012tx, Nepomechie:2013mua}.}
A related question is whether the set of Bethe roots characterizing
any given state must be pairwise distinct, i.e., obey the ``Pauli
principle''.  Although it is generally believed that the answer to
this question is `yes', there is no proof (to our knowledge) of this
assertion \cite{Avdeev:1985cx}.\footnote{The
coordinate Bethe ansatz wavefunction, which is proportional to the
algebraic Bethe ansatz state, does vanish for coincident rapidities.
However, it is only the proportionality factor and not the algebraic
Bethe ansatz state itself that vanishes for coincident rapidities.
The corresponding coordinate Bethe ansatz wavefunction can therefore
be made nonzero by a simple renormalization (of the sort required for
physical singular states), the need for which cannot be precluded.}

We formulate here a precise conjecture for the number of solutions
with pairwise distinct roots of the Bethe equations, in terms of
numbers of singular solutions.  (See Eq.  (\ref{conjecture}) below.)
Its meaning is that the Bethe equations generally have ``too many''
solutions with pairwise distinct roots; but after appropriate culling,
there remain exactly the right number of solutions to account for all
$2^{N}$ eigenstates of the Hamiltonian.

In order to check this conjecture, it is necessary to find {\it all}
solutions with pairwise distinct roots of the Bethe equations.
For certain energy levels, in particular for the ground state which
has $N/2$ real Bethe roots, it is straightforward to compute
numerically the Bethe roots for large values of $N$ ($\sim 10^{3}$)
(see e.g. \cite{Alcaraz:1987zr}), and to compute analytically the
energy in the $N \rightarrow \infty$ limit \cite{Faddeev:1996iy}.
However, finding all solutions of the Bethe equations is unfortunately
a difficult problem even for modest values of $N$ ($\sim 10$) --
certainly much more difficult than directly diagonalizing the Hamiltonian.
For example, on a desktop computer, the direct solution of (\ref{BAE}) is not
feasible for $N=8$ beyond $M=3$.

We discuss two approaches for tackling this problem.  Using
homotopy continuation methods, which heretofore had not been applied to the Bethe
equations,  we find all solutions with pairwise distinct roots up to
$N=14, M=7$.  We also discuss an indirect method of finding
solutions of the Bethe equations by solving the Baxter T-Q equation.

Our results, summarized in Table \ref{table:summary}, are in precise
agreement with the conjecture.  These results also suggest that the
naive prediction (\ref{naive}) for the number of solutions of the
Bethe equations is incorrect not only for small values of $N$, but
also for $N\rightarrow \infty$, in contradiction with several
computations based on the string hypothesis.

The conjecture may also be of interest to mathematicians.  Indeed, the
Bethe equations are evidently a system of polynomial equations, and
therefore belong to the realm of algebraic geometry.  These equations
have a finite number of solutions; i.e., the algebraic variety of the
solutions has dimension 0.  Nevertheless, the number of such solutions
should be calculable a priori.  (See also \cite{Langlands:1995,
Mukhin:2009, Mukhin:2004} and references therein.)

The outline of this paper is as follows. In Section
\ref{sec:completeness} we formulate our conjecture for the number of
solutions with pairwise distinct roots of the Bethe equations, in
terms of numbers of singular solutions. In Section \ref{sec:homotopy}
we briefly describe the  homotopy continuation method with references to some surveys on the method, and how we use it
to solve the Bethe equations. In Section \ref{sec:TQ} we discuss an
indirect way of finding the Bethe roots by solving instead the T-Q
equation. Finally we present a summary and discussion of our results
in Section \ref{sec:summary}.

\section{The completeness/Pauli-principle conjecture}\label{sec:completeness}

For given values of $N$ and $M$, let us denote by ${\cal N}(N,M)$ the
number of solutions of the Bethe equations (\ref{BAE}) with pairwise
distinct Bethe roots (i.e., $\lambda_{j} \ne \lambda_{k}$ for $j\ne
k$).  We always count solutions up to permutations of the Bethe roots: if
$\{\lambda_{1} \,, \ldots \,, \lambda_{M}\}$ is a solution, then any
permutation of these $\lambda_{k}$'s is not counted as a separate solution.

We would like to formulate a conjecture for ${\cal N}(N,M)$.  To this end,
it is necessary to review in a little more detail the solution of the
model.  Instead of following Bethe's original approach (which is now
referred to as the coordinate Bethe ansatz), we find it easier to use
the algebraic Bethe ansatz approach \cite{Korepin:1993,
Faddeev:1996iy, Nepomechie:1998jf}.  The main point is that the state
with all spins up, which is called the ``reference'' state and which
we denote by $|0\rangle$, is an eigenstate of the Hamiltonian (with
eigenvalue 0); and additional eigenstates, called ``Bethe states'',
can be constructed by acting with certain creation operators
$B(\lambda)$ on the reference state:
\be
|\lambda_{1} \,,
\ldots \,, \lambda_{M} \rangle = B(\lambda_{1}) \cdots B(\lambda_{M})
|0\rangle \,,
\label{vec}
\ee
where $\{\lambda_{1} \,, \ldots \,, \lambda_{M}\}$ are (by
assumption) pairwise distinct, and are solutions of the
Bethe equations. Since the creation operators commute
$\left[B(\lambda)\,, B(\lambda') \right]=0$, any permutation of the
Bethe roots $\{\lambda_{1} \,, \ldots \,, \lambda_{M}\}$ evidently does not
affect the state (\ref{vec}). The corresponding energy eigenvalue is given by
(\ref{energy}).

The Hamiltonian (\ref{Hamiltonian}) commutes with the total spin
$\vec S$,
\be
\left[ H \,, \vec S \right] = 0 \,, \qquad \vec S = \frac{1}{2}\sum_{n=1}^{N} \vec
\sigma_{n} \,.
\ee
Hence, $H$, ${\vec S}^{2}$ and $S^{z}$ can all be simultaneously
diagonalized,
\be
H |E, s, m \rangle &=&
E   |E, s, m \rangle \,, \non\\
{\vec S}^{2} |E, s, m \rangle &=&
s(s+1)   |E, s, m \rangle \,, \non\\
S^{z} |E, s, m \rangle &=& m
|E, s, m \rangle \,.
\ee
It can be further shown that the Bethe states (\ref{vec}) are highest-weight
states,
\be
S^{+} |\lambda_{1} \,, \ldots \,, \lambda_{M} \rangle = 0 \,, \qquad
S^{\pm} = S^{x} \pm i S^{y} \,,
\ee
so the spin quantum numbers are given by
\be
s=m=\frac{N}{2}-M \,.
\label{quantumno}
\ee
Although the Bethe ansatz can give only highest-weight ($m=s$)
eigenstates of the Hamiltonian, the eigenstates with $m<s$ can easily
be obtained by repeatedly acting with the spin-lowering operator
$S^{-}$ on the Bethe states.

It is now not difficult to determine, for a given value of $N$, the
possible values of spin $s$ (and hence the possible values of $M$),
and their degeneracies (which naively should be the number of solutions of the
Bethe equations with $M$ Bethe roots).  Indeed, the Clebsch-Gordan
theorem implies that the $N$-fold tensor product of spin-1/2
representations decomposes into a direct sum of (irreducible) spin-$s$
representations,
\be
\underbrace{{\bf \frac{1}{2}}\otimes \cdots \otimes {\bf
\frac{1}{2}}}_{N} =
\bigoplus_{s=0}^{\frac{N}{2}} n_{s} {\bf s}\,,
\label{CG}
\ee
where $n_{s}$, the number of representations with spin-$s$, is given by
\be
n_{s} = {N\choose \frac{N}{2}-s} - {N\choose \frac{N}{2}-s-1} \,.
\label{degeneracy}
\ee
Note from (\ref{CG}) that the values of $s$ range from 0 to $N/2$;
hence, according to (\ref{quantumno}), the values of $M$ also range
from 0 to $N/2$, as already anticipated in (\ref{BAE}).

It follows from (\ref{quantumno}) and  (\ref{degeneracy}) that
${\cal N}(N,M)$, the number of
solutions of the Bethe equations with $M$ pairwise distinct roots, should naively be given by
\be
{\cal N}(N,M) \stackrel{\rm ?}{=} {N\choose M} - {N\choose M-1} \,.
\label{naive}
\ee
However, this is not correct.  The flaw in the argument can be traced
back to the incorrect assumption that every solution of the Bethe
equations with pairwise distinct roots produces, via (\ref{vec}), an eigenstate of the Hamiltonian.
Indeed, the Bethe equations admit so-called singular (or exceptional)
solutions, one of whose roots is $i/2$ and another of which is
$-i/2$; and only a subset of those solutions produces eigenstates of the
Hamiltonian. Those singular solutions that produce eigenstates of the
Hamiltonian we call ``physical'', and those singular solutions that do
not produce eigenstates of the Hamiltonian we call ``unphysical''.

Fortunately, there exists a simple criterion for determining whether
a given singular solution is physical or unphysical. Consider a
general singular solution of the Bethe equations
\be
\left\{ \frac{i}{2}\,, -\frac{i}{2}\,, \lambda_{3}\,, \ldots \,,
\lambda_{M} \right\} \,,
\label{gensing}
\ee
where $\lambda_{3}\,, \ldots \,, \lambda_{M}$ are pairwise distinct
and are not equal to $\pm i/2$.  The Bethe equations
(\ref{BAE}) imply that the last $M-2$ roots $\{ \lambda_{3}\,, \ldots
\,, \lambda_{M} \}$ obey
\be
\left(  \frac{\lambda_{k} + \frac{i}{2}}
{\lambda_{k} - \frac{i}{2}} \right)^{N-1}
\left(\frac{\lambda_{k} - \frac{3i}{2}}
{\lambda_{k} + \frac{3i}{2}}\right)
=  \prod_{\scriptstyle{j \ne k}\atop \scriptstyle{j=3}}^{M}
\frac{\lambda_{k} - \lambda_{j} + i}
{\lambda_{k} - \lambda_{j} - i}
\,, \qquad k = 3 \,, \cdots \,, M \,.
\label{BAE2}
\ee
This singular solution is physical if $\{ \lambda_{3}\,, \ldots
\,, \lambda_{M} \}$ also obey \cite{Nepomechie:2013mua}
\be
\left[-\prod_{k=3}^{M}\left(  \frac{\lambda_{k} + \frac{i}{2}}
{\lambda_{k} - \frac{i}{2}} \right)\right]^{N} =1 \,.
\label{physical}
\ee

We are finally ready to formulate a precise conjecture for ${\cal
N}(N,M)$ in terms of numbers of singular solutions. Let
${\cal N}_{s}(N,M)$ denote the number of solutions of the Bethe
equations (\ref{BAE}) that are singular (i.e., that have the form
(\ref{gensing})); and let ${\cal N}_{sp}(N,M)$
denote the number of such singular solutions that satisfy (\ref{physical})
and hence are physical. We conjecture that
\be
{\cal N}(N,M) -{\cal N}_{s}(N,M) + {\cal N}_{sp}(N,M) = {N\choose M} - {N\choose M-1} \,.
\label{conjecture}
\ee
The physical meaning of this conjecture is that the Bethe equations
generally have ``too many'' solutions with pairwise distinct roots; but
after discarding the singular solutions that do not satisfy
(\ref{physical}), there remain exactly the right number of solutions
to account for all ${N\choose M} - {N\choose M-1}$ highest-weight
eigenstates (and therefore all $2^{N}$ eigenstates) of the
Hamiltonian.  Hence, (\ref{conjecture}) expresses the completeness --
after appropriate culling -- of the solutions of the Bethe equations
with pairwise distinct roots.  It is therefore natural to call it the
``completeness/Pauli-principle conjecture''.\footnote{We could generalize (\ref{conjecture}) to allow
for violations of the Pauli principle by adding an extra term:
\be
{\cal N}(N,M) -{\cal N}_{s}(N,M) + {\cal N}_{sp}(N,M) + {\cal N}_{strange}(N,M) = {N\choose M}
- {N\choose M-1} \,, \non
\ee
where ${\cal N}_{strange}(N,M)$ is the number of ``strange'' solutions of the
Bethe equations whose roots are {\it not} pairwise distinct but which
nevertheless produce, via (\ref{vec}), eigenstates of the Hamiltonian. However, as far as we
have checked, there is no need for such a term, i.e. ${\cal N}_{strange}(N,M)=0$.}
Unfortunately, we do not have conjectures for ${\cal N}\,, {\cal
N}_{s}\,, {\cal N}_{sp}$ separately for general values of $N$ and $M$,
but only for the combination ${\cal N} -{\cal N}_{s} + {\cal N}_{sp}$.

We emphasize the two-way nature of this conjecture:
\renewcommand{\labelenumi}{(\roman{enumi})}
\begin{enumerate}
    \item for every highest-weight eigenstate of the Hamiltonian, there is a solution of
the Bethe equations with pairwise distinct roots; and
    \item for every solution of the Bethe equations with pairwise
    distinct roots that  (if it is singular)
also satisfies (\ref{physical}), there is a highest-weight eigenstate
of the Hamiltonian.
\end{enumerate}
To our knowledge, previous studies of completeness have focused
only on point (i), and therefore have not addressed the many
unphysical singular solutions of the Bethe equations. A case in point
is \cite{Hagemans:2007}, which reports solutions of the Bethe equations for
$N=8$ and $N=10$, but the only singular solutions that are included are the physical ones.

In order to check the conjecture, it is necessary to find {\it all} the
solutions with pairwise distinct roots of the Bethe equations. It is
to this task that we now turn.

\section{Homotopy  continuation}\label{sec:homotopy}

Homotopy continuation (often called continuation) is the main
numerical approach to find isolated roots of polynomial systems
without regard to the dimension of the ambient space, restrictions
on the systems, or the bounds on the size of the domain in Euclidean
space where the solutions are sought. In this section we outline the
method and its main features, and mention the main software packages.

Some surveys of this material are \cite{L97,SW2,BHSW1,WSActaNum11}.
The classic \cite{Mor87} is a good reference for many software
implementation details.  The methods apply more generally than systems
of polynomials, but without many of the features we need, e.g.,
guarantees for finding solutions and accurate computation of singular
solutions.  The classic reference for continuation methods in this
general situation (putting the case of polynomial systems in context)
is \cite{AllgowerGeorg}.

Polynomial system solving has a long history of applications to widely
diverse areas, e.g., kinematics, chemical reaction systems, game
theory, mathematical biology, systems of nonlinear differential
equations, and physics.
Applications to theoretical kinematics go back to the 19th century.
It is fair to say that the main application areas presented in
\cite{Mor87,SW2} lie in kinematics and robotics; for more details, see
the references in those books and the survey article
\cite{WSActaNum11}.
Chemical reactions systems are treated in \cite{Mor87}; for more
details, see the references there and in the articles
\cite{G2001,GB2002,SW2,APPSDHM09}.
Game theory applications and some references may be found in \cite{SW2}.
Polynomial systems arise naturally in the discretization of systems
of nonlinear differential equations;
for more details see the early articles \cite{ABSW06,ACT09a}, some more recent articles,
\cite{HHSSXZ,HHS13}, and the references contain in them.  For applications to a variety of models from
from mathematical biology, see \cite{HHHLSZ1,
HHS12,HHHMS12}.
For recent applications to physics, see e.g. \cite{
Mehta:2011xs, Kastner:2011zz, Mehta:2012qr, Nerattini:2012pi,
MHH,
Maniatis:2013qaa, Hughes:2012hg, Mehta:2011wj,
MartinezPedrera:2012rs,He:2013yk,Greene:2013ida}
and the references contained in these articles.

The continuation method to find the solutions of a polynomial system
$f$ starts with a polynomial system $g$ (called the start system) and
a set ${ \cal S}_g$ of nonsingular solutions of $g$, which are called
start points.  The method proceeds to deform the system $g$ and ${\cal
S}_g$ to the system $f$ and a set of solutions ${\cal S}_f$ of $f$.
We make this precise below.  The references listed above detail many
different possible choices of $g$ for which homotopy continuation is
guaranteed to find a set of solutions of $f$, which contain all
isolated solutions of $f$.  We use the total degree homotopy
(explained below), which is the first homotopy that was shown to find
all solutions of a polynomial system.

For Bethe's equations (\ref{BAE}), we begin by introducing the
polynomials $f_k$ in the variables $\lambda_1,\ldots,\lambda_M$, which we regard as specifying
a variable point in ${\mathbb C}^M$
\be
f_k(\vec{\lambda})=\left(\lambda_{k} + \frac{i}{2}\right)^{N}
\prod_{\scriptstyle{j \ne k}\atop \scriptstyle{j=1}}^M (\lambda_{k}
- \lambda_{j} - i) ~-~ \left(\lambda_{k} - \frac{i}{2}\right)^{N}
\prod_{\scriptstyle{j \ne k}\atop \scriptstyle{j=1}}^M (\lambda_{k}
- \lambda_{j} + i) \,,
\ee
where $\vec{\lambda}=(\lambda_1 \,, \lambda_2\,, \ldots \,, \lambda_M)^T$. Notice that the
degrees the polynomials $f_k$ are $N+M-2$ after simplifying (since the top degree
terms of the LHS and the RHS of Eq. (\ref{BAE}) are equal).

We then define
the following homotopy function
\be
\vec H(\vec{\lambda},t)=(1-t)\vec{f}(\vec{\lambda})+\gamma
t\vec{g}(\vec\lambda),
\ee where
$\vec{f}=(f_1,\cdots,f_M)^T$,
$\vec{g}=(g_1,\cdots,g_M)^T$,
$g_k=\lambda_k^{N+M-2}-1$. Moreover,
$t\in [0, 1]$ is a homotopy parameter, and $\gamma$ is a random
complex number. When $t = 1$, we have known
solutions to $\vec{g}(\vec{\lambda})=0$,
or equivalently, $\vec H(\vec{\lambda},1) = 0$. Specifically, the
solutions of $g_k=0$ are
\be
\lambda_k= \omega^{j_k} \,, \qquad \omega=e^{2\pi i/(N+M-2)} \,,
\qquad j_k=0,1,\cdots,N+M-3 \,.
\label{lambdak}
\ee
The solutions of $\vec{g}(\vec{\lambda})=0$ 
therefore yield values for each $\lambda_k$.
The known solutions are called start points, and
the system $\vec H(\vec{\lambda},1) = 0$ is called the start system.
Such a start system 
with the degree $d_k$ of $g_k$ equal to the degree of $f_k$ for all $k$ and with the number of solutions
of $\vec H(\vec{\lambda},1) = 0$ equal to $\prod_{k=1}^n d_k$  is called a {\em total
degree start system}. (The number of isolated solution of a system
of $n$ polynomial equations $f_k$ in $n$ variables is always less than
or equal to $\prod_{i=k}^n \deg{f_k}$.) Choosing a total degree start system
and a random complex number $\gamma$ guarantees finding all the solutions.  The use of the random $\gamma$, which is
called the $\gamma$-trick, was introduced in \cite{MS87}.  A good discussion of a more general version of this trick is
given in \cite[Lemma 7.1.3]{SW2}.

However, as previously
explained, we are interested in solutions $(\lambda_1, \cdots,
\lambda_M)$ up to permuting the coordinates, and with no two
$\lambda$'s equal. Hence, we may restrict $j_k$ in (\ref{lambdak})
to run over all $M$-tuples of integers \be 0\le
j_1<j_2<j_3<\cdots<j_M \le N+M-3 \,. \label{ranges} \ee To see this,
note that the equations $H_1,\ldots,H_M$ of the homotopy $\vec
H(\vec{\lambda},t)$ are permuted by the symmetric group on the
variables $\lambda_1,\ldots, \lambda_M$.  Indeed under a permutation
$\sigma$ taking $\lambda_j$ to $\lambda_k$, $H_j$ is taken to $H_k$.
This has strong consequences for the homotopy \cite{VC1994}: we
explain the consequence we need. The paths in $\lambda,t$ space over
the interval $(0,1]$, are permuted by the action of the symmetric
group. A start point has pairwise disjoint entries if and only if
the orbit in the set of start points under the symmetric group
consists of exactly $M!$ points. Since there is one path for each of
the start points, we see that the number of paths a given path over
$(0,1]$ is taken to under the symmetric group is $M!$ if and only if
the start point has pairwise disjoint entries. The orbit of a root
of $\vec H(\vec{\lambda},0) = 0$, which is a limit of a path $p$,
equals the set of limits as $t\to 0$ of the paths in the orbit of
the path $p$ under the symmetric group action. From this we see that
a root of $\vec H(\vec{\lambda},0) = 0$ can have pairwise disjoint
entries only if it is the limit of some path with a start point
having pairwise disjoint entries.  Note that this does not preclude
a path with start point having pairwise disjoint coordinates ending
at a root of  $\vec H(\vec{\lambda},0) = 0$ without pairwise
disjoint coordinates.
Note also that the number of start points is only ${N+M-2\choose M}$ since it
equals the number of $M$-tuples of integers (\ref{ranges}).

At $t = 0$, we evidently recover the
Bethe equations. The problem of finding the solutions of the Bethe
equations now reduces to tracking solutions of $\vec
H(\vec{\lambda},t) = 0$ from $t = 1$ where we know solutions to $t =
0$. The numerical method used in path tracking from $t = 1$ to $t =
0$ arises from solving the {\em Davidenko differential equation:}
\bes \frac{d \vec H(\vec{\lambda}(t),t)}{dt}=\frac{\partial \vec
H(\vec{\lambda}(t),t)}{\partial
\vec{\lambda}}\frac{d\vec{\lambda}(t)}{dt}+\frac{\partial \vec
H(\vec{\lambda}(t),t)}{\partial t}=0 \,. \label{TDH}\ees In
particular, path tracking reduces to solving initial value problems
numerically with the start points being the initial conditions.
Since we also have an equation which vanishes along the path, namely
$\vec H(\vec{\lambda},t) = 0$, predictor/corrector methods, e.g.,
rkf45 as a predictor with the Newton-Rapfson's method as a corrector, are used 
to solve these initial value problems.

Continuation methods parallelize naturally, by sending different paths to different processors to track.
Predictor/corrector methods combined with adaptive stepsize and
adaptive precision algorithms \cite{BHSW2,BHSW3} provide reliability
without giving up efficiency.
A major
concerns for implementing a numerical path-tracking algorithm are to
decide the number of digits used to provide reliable computation,
which predictor/corrector method to employ and the stepsize $\Delta
t$. See \cite{BHSW3,SW2} for more details regarding the construction
and implementation of a path tracking algorithm.

We mention that for polynomial systems, there are special path-tracking algorithms  \cite{MSW91,MSW92a,MSW92b}, often called endgames, to
compute singular solutions.   When the endpoint of a solution path is
singular, there are several approaches that give highly accurate estimates of the endpoint. These methods
use the fact that the homotopy
continuation path $\vec{\lambda}(t)$ approaching a solution of $\vec
H(\vec{\lambda},t) = 0$ as $t\rightarrow 0$ lies on an complex
analytic curve, which may be locally uniformized near $(\vec{\lambda},0)$,  by an analytic disk. Many of these endgames  are
implemented in several sophisticated numerical packages
such as Bertini \cite{BHSW1},
PHCpack \cite{V},  and HOMPACK \cite{WSMMW}. Their binaries are all
available as freeware from their respective research groups.  We note also
HOM4PS-2.0 \cite{LLT}, the leading homotopy continuation software using polyhedral methods: this software
is not useful for us in this article because it does not allow the user to specify a homotopy and a set of start points.

We employed Bertini \cite{BHSW1} with adaptive precision tracking
\cite{BHSW2,BHSW3}: \{due to the high degree of Bethe's equations, a
precision of 2000 bits (roughly 600 digits)\} is needed to solve the
system. We ran Bertini on a 64-bit Linux cluster with 13 dual Xeon
5410 nodes (each with 8 GB RAM and 8 cores) and 9 dual Xeon E5520
nodes (each with 12 GB RAM and 8 cores). One node acted as manager
with up to 22 computing nodes giving a total of 176 cores).

The results for $N=2\,, \ldots \,, 12$ are presented in a set of
supplemental tables \cite{tables}, an example of which is Table \ref{table:N8M4}.
To give some perspective in computational effort, the case of $(N,M)=(14,7)$
took 4413.5 seconds of computation while the case of $(N,M)=(12,6)$ took 168.8 seconds of computation.

\begin{table}[htb]
  \tiny
  \centering
  \begin{tabular}{|c| c c c|}\hline
      number & &  Bethe roots $\{ \lambda_{k} \}$ &   \\
    \hline
1& $\pm$ 0.5250121022236669 &
$\pm$ 0.1294729463749287 &   \\
2&0.5570702385744416 &0.1470126111961413 &
-0.3520414248852914 $\pm$ 0.5005581696433306I  \\
3*&$\pm$ 0.5I& -0.2930497652740115 $\pm$
0.5002695484553508I &  \\
4*&$\pm$ 0.5I& 0.09053461122303935 & 0.4866819617430914 \\
5*&$\pm$ 0.5I&-0.04929340793103601+1.631134975618312I&
-0.2430919428911911-0.06188079036780695I  \\
6*&$\pm$ 0.5I &
0.6439488581706157-0.1197616885579488I&0.05986712277687283+1.57171694471433I  \\
7*&$\pm$ 0.5I & 0.04929340793103601+1.631134975618312I
&0.2430919428911911-0.06188079036780695I \\
8**&$\pm$ 0.5I & $\pm$ 0.5638252623934961 & \\
9&0.2205600072920844 &-0.6691229228815117 &
0.2242814577947136 $\pm$ 1.002247276506607I  \\
10*&$\pm$ 0.5I & 0.1695810016454493 &-0.522716443014433  \\
11*&$\pm$ 0.5I &-0.05986712277687283+1.57171694471433I
&-0.6439488581706157-0.1197616885579488I \\
12*&$\pm$ 0.5I & 1.653144833689466I &-0.050307293346599I \\
13*&$\pm$ 0.5I & 0.522716443014433  &-0.1695810016454493  \\
14*&$\pm$ 0.5I &0.050307293346599I &-1.653144833689466I \\
15*&$\pm$ 0.5I &-0.4866819617430914 &-0.09053461122303935  \\
16*&$\pm$ 0.5I & 0.04929340793103601-1.631134975618312I &
0.2430919428911911+0.06188079036780695I \\
17*&$\pm$ 0.5I & 0.2930497652740115 $\pm$ 0.5002695484553508I &  \\
18**&$\pm$ 0.5I & $\pm$ 0.1424690678305666 &  \\
19**&$\pm$ 0.5I & $\pm$ 1.556126503577051I &  \\
20*&$\pm$ 0.5I & 3.517084291308099I&1.508105736964082I  \\
21&0.2443331937711654 &-0.08378710739142802 & -0.08027304318986867 $\pm$
1.005588273959932I \\
22*&$\pm$ 0.5I & 0.05986712277687283-1.57171694471433I &
0.6439488581706157+0.1197616885579488I  \\
23& 0.1211861779691729
&-0.5716111771864383 & 0.2252124996086327 $\pm$ 0.5000288621635332I  \\
24&  & $\pm$ 0.4632647275890309 $\pm$
0.5022938535699026I &   \\
25& -0.1470126111961413 &-0.5570702385744416 & 0.3520414248852914
$\pm$ 0.5005581696433306I  \\
26*&$\pm$ 0.5I & -0.05986712277687283-1.57171694471433I
&-0.6439488581706157+0.1197616885579488I  \\
27& 0.08378710739142802 &-0.2443331937711654 & 0.08027304318986867
$\pm$ 1.005588273959932I  \\
28*&$\pm$ 0.5I &-0.2430919428911911+0.06188079036780695I
&-0.04929340793103601-1.631134975618312I  \\
29& $\pm$ 1.025705081230743I & $\pm$ 0.0413091275245562 & \\
30*&$\pm$ 0.5I &-1.508105736964082I&-3.517084291308099I  \\
31& 0.5716111771864383 &-0.1211861779691729 & -0.2252124996086327
$\pm$ 0.5000288621635332I  \\
32&-0.2205600072920844 &0.6691229228815117 & -0.2242814577947136
$\pm$ 1.002247276506607I  \\
\hline
   \end{tabular}
\caption{\small Solutions with distinct roots of the Bethe equations
for $N=8, M=4$.  Singular solutions that are unphysical are labeled by $*$, and singular
solutions that are physical are labeled by $**$.}
   \label{table:N8M4}
\end{table}

\section{Solving the T-Q equation}\label{sec:TQ}

There is also a powerful indirect way of determining the Bethe roots based
on the Baxter T-Q equation. By virtue of the model's integrability,
one can construct the so-called transfer matrix $t(\lambda)$,
a $2^{N} \times 2^{N}$ matrix that is a function of an arbitrary
parameter $\lambda$, which commutes with itself for different values of
this parameter as well as with the Hamiltonian (\ref{Hamiltonian}) \cite{Baxter1982,
Korepin:1993, Faddeev:1996iy, Nepomechie:1998jf}:
\be
\left[ t(\lambda) \,, t(\lambda') \right] = 0 \,, \qquad
\left[ t(\lambda) \,, H \right] = 0 \,.
\ee
Furthermore, it can be shown that the eigenvalues of the transfer
matrix, which we denote here by $T(\lambda)$, are polynomials in
$\lambda$ of degree $N$,
\be
T(\lambda) = \sum_{j=0}^{N}T_{j} \lambda^{j} \,,
\label{T}
\ee
where the coefficients $T_{j}$ are independent of $\lambda$.
Moreover, the transfer matrix eigenvalues satisfy the celebrated T-Q equation
\cite{Baxter1982, Korepin:1993, Faddeev:1996iy, Nepomechie:1998jf}
\be
T(\lambda)\, Q(\lambda) = \left(\lambda+\frac{i}{2}\right)^{N} Q(\lambda-i) +
\left(\lambda-\frac{i}{2}\right)^{N} Q(\lambda+i) \,,
\label{TQ}
\ee
where $Q(\lambda)$ is a polynomial in $\lambda$ of degree $M$, whose
zeros are the sought-after solutions of the Bethe equations (\ref{BAE})
\be
Q(\lambda) = \prod_{m=1}^{M} (\lambda-\lambda_{m}) =
\sum_{j=0}^{M}Q_{j} \lambda^{j} \,, \qquad Q_{M} = 1 \,,
\label{Q}
\ee
where the coefficients $Q_{j}$ are independent of $\lambda$. Indeed,
dividing both sides of (\ref{TQ}) by $Q(\lambda)$, it appears that
the RHS has poles at the zeros of $Q(\lambda)$, in contradiction with the fact the LHS is a
polynomial in $\lambda$ and must therefore be regular. The only
way out of this paradox is for the poles to cancel, the condition for which is
precisely the Bethe equations (\ref{BAE}).

Interestingly, it is possible to solve the T-Q equation (\ref{TQ})
numerically for both $T(\lambda)$ and $Q(\lambda)$; and then, by finding the zeros of
$Q(\lambda)$, determine all the solutions of the Bethe
equations.\footnote{This very nice approach does not seem to be
widely known. To our knowledge, it was first published
by Baxter \cite{Baxter:2001sx}, and it has recently been rediscovered in the
context of Richardson-Gaudin models \cite{Pan:2011, Marquette:2012}.}
The basic idea is to substitute (\ref{T}) and (\ref{Q}) into the T-Q
equation (\ref{TQ}), and then equate coefficients with equal
powers of $\lambda$. In other words, (\ref{T}) and (\ref{Q}) imply
that
\be
T(\lambda)\, Q(\lambda) -\left[ \left(\lambda+\frac{i}{2}\right)^{N} Q(\lambda-i) +
\left(\lambda-\frac{i}{2}\right)^{N} Q(\lambda+i) \right] = \sum_{j=0}^{N+M}
c_{j} \lambda^{j} \,,
\ee
where the coefficients $c_{j}$ are independent of $\lambda$; and the T-Q
equation implies that
\be
c_{j} = 0 \,, \qquad j=0\,, 1 \,, \ldots\,, N+M \,.
\label{cj}
\ee
Note that $c_{M+1}\,, \ldots \,, c_{M+N}$ are independent of $T_{0}$
and are linear in $T_{1}\,, \ldots\,, T_{N}$. Therefore, Eqs. (\ref{cj})
with $j=M+1\,, \ldots\,, M+N$ can be solved uniquely for $T_{1}\,,
\ldots\,, T_{N}$ in terms of $Q_{0}\,, \ldots\,, Q_{M-1}$.
Substituting this solution into the remaining coefficients $c_{0}\,,
\ldots \,, c_{M}$, we arrive at a system of $M+1$ nonlinear equations
\be
c_{j} = 0 \,, \qquad j=0\,, 1 \,, \ldots\,, M
\label{cj2}
\ee
for the $M+1$ unknowns $Q_{0}\,, \ldots\,, Q_{M-1}\,,
T_{0}$.\footnote{Since $c_{0}\,, \ldots \,, c_{M}$ are linear in
$T_{0}$, it is also possible to eliminate $T_{0}$ in terms of $Q_{0}\,,
\ldots\,, Q_{M-1}$, and thereby arrive at a system of $M$ nonlinear
equations for $M$ unknowns.} This system of equations is somewhat simpler than
the corresponding $M$ Bethe equations (\ref{BAE}). Indeed, the above procedure can
easily be implemented on a desktop computer, which can perform the
case $N=9$ in a few minutes, but cannot manage higher values of $N$.
The results agree with the corresponding results (up to $N=9$) obtained in
Section \ref{sec:homotopy}. We have not attempted to implement this
procedure on better hardware.

A variation of the above procedure is to first determine the
eigenvalues $T(\lambda)$ by explicitly diagonalizing the transfer
matrix $t(\lambda)$ \footnote{In practice, one first computes the
eigenvectors of the numerical matrix $t(\lambda_{0})$ (where
$\lambda_{0}$ is some generic numerical value), and then one acts with
$t(\lambda)$ on these eigenvectors to obtain the corresponding
eigenvalues $T(\lambda)$ as polynomials in $\lambda$.} and then
solving the T-Q equation (\ref{TQ}) for $Q(\lambda)$.\footnote{This
approach for determining Bethe roots has been used many times in the
past, e.g. \cite{Fabricius:2000yx, Albertini:1991pf, Nepomechie:2003ez}.}
This approach has the advantage that solving the T-Q
equation for only Q is a linear (albeit, overdetermined) problem;
however, it has the disadvantage of requiring the diagonalization of a
large matrix.  This procedure can also be easily implemented on a
desktop computer, which can again perform the case $N=9$ in a few
minutes, but cannot manage higher values of $N$. Unlike the former approach
where one solves the T-Q equation for both T and Q, the only singular
solutions that this method can generate are the physical ones. Hence,
this method can check only point (i) of the two points listed below (\ref{conjecture}).

\section{Summary and Discussion}\label{sec:summary}

Our results are summarized in Table \ref{table:summary}. For each set of values
$(N,M)$, we report a set of four integers:
\be
({\cal N}\,, {\cal N}_{s}\,, {\cal N}_{sp}\,;
{\cal N}-{\cal N}_{s}+{\cal N}_{sp}) \,, \non
\ee
where ${\cal N}$ is the number of solutions with
pairwise distinct roots of the Bethe equations; ${\cal N}_{s}$ is the
number of singular solutions; and  ${\cal N}_{sp}$ is the number of
singular solutions that are physical. (See Section
\ref{sec:completeness} for further details.) These quantities can
easily be read off from
Table \ref{table:N8M4} and the supplemental tables \cite{tables} as follows: ${\cal N}$ is the number of solutions
listed in a given table; ${\cal N}_{s}$ is the number of those
solutions labeled with either a single $*$ or double $**$ star; and ${\cal N}_{sp}$ is the number of those
solutions labeled with a double star.

\begin{table}[htb]
    \tiny
  \centering
  \begin{tabular}{|l|c|c|c|c|c|c|c|}\hline
    \diagbox[width=2.5em]{$N$}{$M$} & 1 & 2 & 3 & 4 & 5 & 6 & 7 \\
    \hline
     2 & (1,0,0; 1) & & & & & &\\
     3 & (2,0,0; 2) & & & & & &\\
     4 & (3,0,0; 3) & (2, 1, 1; 2) & & & & &\\
     5 & (4,0,0; 4) & (6, 1, 0; 5) & & & & &\\
     6 & (5,0,0; 5) & (9, 1, 1; 9) & (9, 5, 1; 5) & & & &\\
     7 & (6,0,0; 6) & (15, 1, 0; 14) & (20, 6, 0; 14) & & & &\\
     8 & (7,0,0; 7) & (20, 1, 1; 20) & (34, 7, 1; 28) & (32, 21, 3;
     14) & & &\\
     9 & (8,0,0; 8) & (28, 1, 0; 27) & (54, 8, 2; 48) & (69, 27, 0;
     42) & & &\\
     10& (9,0,0; 9) & (35, 1, 1; 35) & (83, 9, 1; 75) & (122, 36, 4;
     90) & (122, 84, 4; 42) & &\\
     11& (10,0,0; 10) & (45, 1, 0; 44) & (120, 10, 0; 110) & (209,
     44, 0; 165) & (252, 120, 0; 132) & &\\
     12& (11,0,0; 11) & (54, 1, 1; 54) & (163, 10, 1; 154) & (325,
     55, 5; 275) & (456, 163, 4; 297) & (452, 330, 10; 132) & \\
     13& (12, 0, 0; 12) & (66, 1, 0; 65) & (220, 12, 0; 208) & (494,
     65, 0; 429) & (792, 220, 0; 572) & (919, 490, 0, 429) & \\
     14& (13, 0, 0; 13) & (77, 1, 1; 77) & (285, 13, 1; 273) & (709,
     78, 6; 637) & (1281, 286, 6; 1001) & (1701, 715, 15; 1001) &
     (1701, 1287, 15; 429) \\
    \hline
   \end{tabular}
   \caption{\small The values $({\cal N}\,, {\cal N}_{s}\,, {\cal N}_{sp}\,;
{\cal N}-{\cal N}_{s}+{\cal N}_{sp})$ for given values
of $N$ and $M$, where ${\cal N}$ is the number of solutions with
pairwise distinct roots of the Bethe equations; ${\cal N}_{s}$ is the
number of singular solutions; and  ${\cal N}_{sp}$ is the number of
singular solutions that are physical.}
   \label{table:summary}
\end{table}

Remarkably, the quantities ${\cal N}-{\cal N}_{s}+{\cal N}_{sp}$ in
all the entries of Table \ref{table:summary} coincide with ${N\choose
M} - {N\choose M-1}$, in perfect agreement with the conjecture
(\ref{conjecture}).  Although this conjecture was motivated from consideration
of a physical model (\ref{Hamiltonian}), it can be viewed solely as a
statement about the solutions of the polynomial equations (\ref{BAE})
and (\ref{physical}), which begs for a proof.

It is easy to see that the number of solutions for $M=1$ is $N-1$,
\be
{\cal N}(N,1) = N-1 \,.
\ee
Moreover, for $M=2$, we observe
\be
{\cal N}(N,2) = \frac{1}{2}\left(N^{2}+3N+1+(-1)^{N} \right) \,.
\ee
It would be interesting to formulate conjectures (not to mention
proofs) for ${\cal N}(N,M)$ for $M
\ge 3$.

Several remarks about the singular solutions are in order:
\begin{enumerate}

    \item Inspection of Table \ref{table:N8M4} and the supplemental
    tables \cite{tables} shows that many (but not all) of the
    unphysical singular solutions (i.e., those solutions labeled by a
    single star $*$) are not self-conjugate.  This does not violate
    any theorems, since only solutions corresponding to eigenstates of
    the Hamiltonian are required to be invariant under complex
    conjugation \cite{Vladimirov:1986}.  Such solutions definitely do
    {\it not} obey the string hypothesis, since string configurations
    are (by definition) self-conjugate.

    \item For odd values of $N$, it appears from Table
    \ref{table:summary} that most singular solutions are
    unphysical; i.e., ${\cal N}_{sp}(N,M)=0$ for most
    values of $M$ if $N$ is odd. An exception is the case $N=9, M=3$, for which
    ${\cal N}_{sp}(9,3)=2$; and this repeats with a periodicity of 6:
    ${\cal N}_{sp}(15,3)=2$, etc. We expect that similar
    exceptions occur for higher values of $M$.

    \item It appears from Table
    \ref{table:summary} that, among the singular solutions,
    relatively few are physical (i.e., generally ${\cal N}_{s}\gg {\cal
    N}_{sp}$), with ${\cal N}_{sp} \sim M$.

   \item For $M \sim N/2$ and most values of $N$, it appears from Table
    \ref{table:summary} that the number of
   unphysical singular solutions ${\cal N}_{s} - {\cal N}_{sp}$ is
   comparable to the number of highest-weight states of the
   Hamiltonian.  This suggests that the naive formula (\ref{naive})
   for the number of solutions of the Bethe equations is incorrect not
   only for small values of $N$, but also for $N\rightarrow \infty$.
   This, in turn, suggests that the computations claiming to prove
   this naive formula for $N\rightarrow \infty$ \cite{Faddeev:1996iy,
   Bethe:1931hc, Takahashi1971, Kirillov:1985} are also incorrect.
   These computations nevertheless manage to obtain the correct number
   of highest-weight states, perhaps because the assumption of the
   string hypothesis (which all of these computations make)
   effectively projects out sufficiently many of these unphysical
   singular solutions (many of which, as noted above, do not obey the
   string hypothesis) in the thermodynamic limit.  It would be
   interesting to understand this point better, given that the
   criterion for selecting the physical singular solutions is not that
   they obey the string hypothesis, but instead (\ref{physical}).

    \end{enumerate}

We have seen that, using appropriate computer resources, numerical
homotopy methods are effective for solving the Bethe equations
directly.  It would be interesting to also implement the indirect
approach based on the T-Q equation using similar computer resources,
and compare the effectiveness of these approaches.

As noted in the Introduction, the model that we have studied here is
the prototype of an entire zoo of quantum integrable models, which
have generalizations of the Bethe equations.  We expect that the
methods used here can also be used to study completeness in some of
these other models.

\section*{Acknowledgments}
We are indebted to Wolfram Decker for his collaboration at an early
stage of this project, and for bringing the authors into contact.
The work of RN was supported in part by the National Science
Foundation under Grants PHY-0854366 and PHY-1212337, and by a Cooper
fellowship. WH and AS would like to thank DARPA/AFRL Grant G-2457-2
for supporting their work.


\providecommand{\href}[2]{#2}\begingroup\raggedright\endgroup

\end{document}